\documentclass{llncs}
\usepackage{makeidx}  % allows for indexgeneration
\usepackage{algorithm}
\usepackage[noend]{algpseudocode}
\usepackage{graphicx}
\usepackage{hyperref}
\graphicspath{ {images/} }
\usepackage{booktabs}
\usepackage[spanish]{babel}
\usepackage[utf8]{inputenc}
\begin{document}
\frontmatter          % for the preliminaries
\pagestyle{headings}  % switches on printing of running heads
\addtocmark{Hamiltonian Mechanics} % additional mark in the TOC

\mainmatter              % start of the contributions
\title{Análisis de resultados electorales: comparación de métodos y estimación de votos por franja etaria}
\titlerunning{Análisis de resultados electorales}  % abbreviated title (for running head)

%Camilo Melani1, Joaquín Torre Zaffaroni1, Alejandro Hernandez1, Juan Echagüe1
\author{Camilo Melani, Joaquín Torré Zaffaroni, Alejandro Hernandez, Juan Echagüe}
\institute{Pragma Consultores, San Martín 550 | (C1004AAL) Buenos Aires - Argentina
\email{\{cmelani,jtorre,alhernandez,jechague\}@pragmaconsultores.com}}
%Jeffrey Dean \and David Grove \and Craig Chambers \and %Kim~B.~Bruce \and
%Elsa Bertino}
%
%\authorrunning{Ivar Ekeland et al.} % abbreviated author list (for running head)
%Craig Chambers, Kim B. Bruce, and Elisa Bertino}
%
%\institute{Princeton University, Princeton NJ 08544, USA,\\
%\email{I.Ekeland@princeton.edu},\\ WWW home page:
%\texttt{http://users/\homedir iekeland/web/welcome.html}
%\and
%Universit\'{e} de Paris-Sud,
%Laboratoire d'Analyse Num\'{e}rique, B\^{a}timent 425,\\
%F-91405 Orsay Cedex, France}

\pagestyle{empty}
\maketitle              % typeset the title of the contribution

\begin{abstract}
El Estado, como organizador de la Sociedad, y sus ciudadanos generan enormes cantidades de datos. Éstos, una vez almacenados y procesados, pueden dar soporte a inquietudes de ramas de las Ciencias Sociales, donde es frecuente encontrarnos con el problema de razonar sobre unidades de análisis diferentes a aquellas sobre las que tenemos datos. En las votaciones, podríamos querer entender la decisión de cada elector o de grupos de electores, cuando inicialmente disponemos sólo de información sobre los resultados agrupados por mesa. El problema de estimar valores para una unidad utilizando valores de la otra se conoce como inferencia ecológica. En este trabajo, comparamos dos métodos para estimar el porcentaje de votos para cada partido, en base a la franja etaria del electorado. El trabajo se complementa con análisis de variación entre votos de primera y de segunda vuelta, y de un plebiscito independiente durante la primera vuelta. Realizamos nuestro análisis utilizando resultados de elecciones nacionales de 2014 en Uruguay.
\end{abstract}
\section{Introducción}
Frecuentemente nos encontramos buscando analizar información desde un punto de vista diferente al cual se puede deducir de manera directa a partir de los datos de los que disponemos. En los resultados electorales, podríamos querer estimar los votos a partir de un agrupamiento de individuos y analizar su evolución en el tiempo, de manera tal de poder tomar decisiones en base a esto. Por ejemplo, si un partido determina que un grupo específico de la comunidad no acompañó sus propuestas, podría promover acciones, leyes u obras que beneficien a dicho grupo, de manera de obtener más votos en la siguiente elección.

En el proyecto discutido en el presente trabajo, nos propusimos analizar los resultados de las elecciones nacionales de 2014 en Uruguay~\cite{OscarJofre}\cite{Penillanura}. Más precisamente, nos propusimos dividir el electorado en franjas etarias y analizar por ende el apoyo de la población hacia tal o cual partido político. Realizamos el análisis utilizando dos métodos diferentes, uno de promedios ponderados de probabilidades y otro tradicional de inferencia ecológica~\cite{GoodmanEcologicalRegressions}\cite{VentajasLimitacionesIE}; llegamos a conclusiones muy similares, lo cual por un lado reafirma la precisión de los métodos, y por el otro nos da herramientas para confiar en los resultados obtenidos. Complementariamente, realizamos diversos análisis adicionales utilizando el método de King. Se pudo ver la evolución del voto de la primera a la segunda vuelta electoral, particularmente sobre cómo se redirigieron los votos en el ballotage. También pudimos analizar cómo se subdividieron los votos de las diferentes listas de diputados presentadas por los partidos que compitieron en la primera vuelta. Por último, analizamos los resultados de un plebiscito sobre reducción de la edad de imputabilidad (de 18 a 16 años) realizado en paralelo con la primera vuelta electoral, para el cual pudimos estimar qué pensarían los votantes de cada partido, en cuanto a esa pregunta en particular.

En la Sección \ref{sec:sistemaelectoraluruguay} presentamos el sistema electoral de Uruguay, y algunas condiciones y supuestos importantes para nuestro trabajo. En la Sección \ref{sec:analisisrealizados} mostramos los resultados de la estimación de voto por franja etaria con los dos métodos seleccionados, e introducimos los análisis adicionales realizados. Finalmente, en la Sección \ref{sec:conclusion} concluimos.

\section{El sistema electoral de Uruguay}
\label{sec:sistemaelectoraluruguay}
Nos propusimos analizar los datos de las elecciones presidenciales de la República Oriental del Uruguay realizadas en Octubre y Noviembre 2014. El objetivo es lograr analizar diferentes interrogantes sobre el comportamiento de los electores, utilizando exclusivamente los datos agregados, puestos a disposición por las autoridades locales.
Uruguay tiene un proceso electoral elaborado que se extiende a lo largo de cerca de un año. Este proceso incluye hasta 4 actos electorales correspondientes a elecciones primarias simultáneas en junio; elecciones legislativas y presidenciales en un mismo acto, en el que también se realizan plebiscitos, en el mes de octubre; una eventual segunda vuelta para las presidenciales en noviembre, y elección de autoridades locales para los 19 departamentos y varias decenas de alcaldías en mayo del año siguiente.

En el proceso electoral iniciado en 2014 hay 2.620.791 personas habilitadas para votar. Para ello cada elector está identificado por un documento llamado Credencial Cívica, que sirve exclusivamente a fines electorales. La Credencial Cívica incluye una serie y un número: la serie se asigna de acuerdo al lugar de residencia, y el número de manera consecutiva dentro de la serie.

Las mesas de votación incluyen alrededor de 400 electores que pertenecen a la misma serie (por ende tienen domicilios cercanos) y con números consecutivos (por ende con edades cercanas). Por ello cada mesa no es una muestra aleatoria de la población, sino que tiene un fuerte sesgo territorial y etario; mientras que el conjunto de mesas para una misma serie suele representar el territorio e incluye votantes de todas las edades. Los datos que utilizamos en este trabajo son los resultados de los distintos actos electorales en cada mesa del país, y el padrón electoral, que indica la edad de cada elector.

Para nuestro análisis, asumimos que la distribución de probabilidades para una determinada franja etaria es uniforme. Esto significa que los votantes de edad similar votan de manera similar, independientemente de otras variables como el territorio en donde habitan o la clase social a la que pertenecen.

\section{Análisis realizados}
\label{sec:analisisrealizados}

Utilizamos dos métodos para estimar los parámetros de las funciones de probabilidad de las distribuciones condicionales: un modelo de promedio ponderado de distribuciones multinomiales, el cual utiliza probabilidades para realizar las estimaciones; y un modelo multinomial-dirichlet jerárquico, el cual utiliza tablas RxC~\cite{KosukeCommonFramework}\cite{KingLibro}\cite{BayesianRxC}. Utilizamos a su vez el paquete para R llamado Zelig~\cite{Zelig} para realizar el procedimiento de inferencia ecológica.

\begin{figure}
    \centering
    \includegraphics[width=12cm,height=6cm]{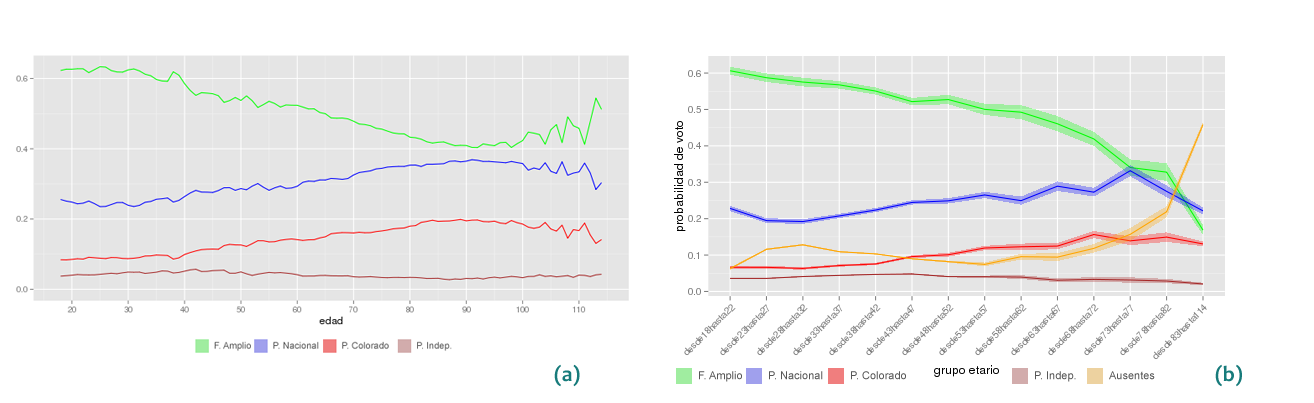}
  \caption{Probabilidad de voto a partido según franja etaria.}
  \label{fig:correlacionPrimeraVuelta}
\end{figure}

Las soluciones implementadas nos permitieron comprender la relación entre la edad y los resultados de la elección general a presidente en la primera vuelta. En la Figura \ref{fig:correlacionPrimeraVuelta} vemos sendos gráficos que nos muestran la probabilidad de que un elector de una franja etaria dada, vote por un determinado partido. Es interesante ver que ambos gráficos son cualitativamente similares.

La Figura \ref{fig:correlacionPrimeraVuelta}.a se construyó con franjas etarias de una única edad, es decir cada edad en una franja diferente, y abarca el electorado del distrito Montevideo. Fue realizado utilizando el modelo de promedio ponderado, el cual para realizar los cálculos toma como probabilidad objetivo la distribución de personas en el padrón de cada mesa, y luego realiza un promedio ponderado de los resultados intermedios.

La Figura \ref{fig:correlacionPrimeraVuelta}.b se construyó para franjas etarias de amplitud 5 años, y también abarca el electorado del distrito Montevideo. Fue realizado utilizando el modelo multinomial-dirichlet jerárquico, el cual para realizar los cálculos toma una matriz con marginales conocidos (los votos resultantes por mesa y la cantidad de personas por edad) y luego de varias simulaciones obtiene los componentes de la matriz. Este método permite también estimar la desviación estándar para cada franja etaria seleccionada, y se muestra en el gráfico con un sombreado alrededor de la curva. En este gráfico también se incluyó una curva para estimar la probabilidad de ausencia en cada franja etaria.

\subsection{Análisis complementarios}

Luego de realizar la comparación entre métodos restringiendo al distrito Montevideo, y comprobar que ambos métodos producen resultados similares, nos propusimos analizar con mayor detalle el electorado completo, estudiando a su vez otras características de interés. También realizamos verificaciones de los modelos, utilizando datos de aprendizaje y de test.

\begin{figure}
  \centering
    %la original es de 421x477
    \includegraphics[width=7cm,height=8cm]{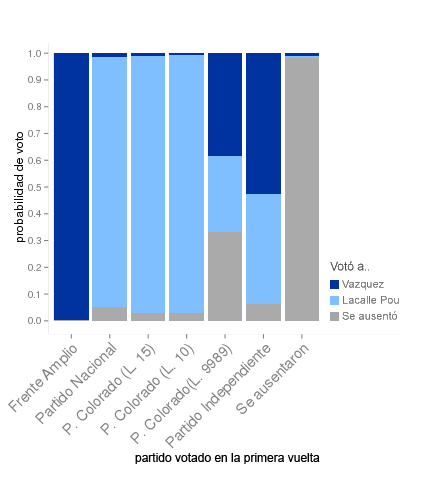}
  \caption{Correlación entre voto en primera vuelta y ballotage,}
  \label{fig:correlacionBallotage}
\end{figure}

Un análisis de probabilidades condicionadas nos permitió estimar cómo se distribuyeron, luego de la primera vuelta, los votantes de los partidos que no compitieron en segunda vuelta. En la Figura \ref{fig:correlacionBallotage} vemos el resultado de este análisis. Podemos determinar que los votantes del Partido Colorado se volcaron por el Partido Nacional, lo cual se alinea con las expresiones de apoyo que realizaron los candidatos que quedaron afuera en primera vuelta.

\begin{figure}
  \centering
    \includegraphics[width=8.5cm,height=7cm]{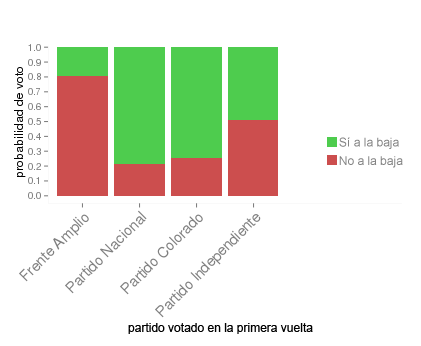}
  \caption{Correlación entre voto a partido y voto en plebiscito.}
  \label{fig:correlacionPlebiscito}
\end{figure}

Otro análisis interesante fue entender qué habían votado en un plebiscito los votantes de los diferentes partidos que compitieron en primera vuelta. Utilizando el método de inferencia ecológica, en este caso definiendo como marginales a los resultados de la primera vuelta y al ``sí a la baja de la edad de imputabilidad'', logramos determinar la probabilidad de que un votante de un partido haya votado sí en el plebiscito. En la Figura \ref{fig:correlacionPlebiscito} podemos observar que los votantes del Frente Amplio estuvieron en su gran mayoría en contra de bajar la edad de imputabilidad. En cambio, los votantes del resto de los partidos se mostraron a favor de bajar la edad.

En todos estos análisis, nuestras estimaciones resultaron altamente correlacionadas a diferentes encuestas de diversos diarios y medios especializados~\cite{diarioEncuestas}. En nuestro caso, no necesitamos prácticamente conocimiento del dominio, sino que realizamos las estimaciones puramente operando con los datos, lo cual es una característica típica de tareas de Big Data.

\section{Conclusión}
\label{sec:conclusion}
En este trabajo, mostramos una amplia gama de resultados relacionados con analizar el comportamiento de los votantes, dados solamente los datos públicos disponibles, logrando desagruparlos e inferir información más detallada. Existen técnicas para llevar adelante estas tareas, aunque también propusimos una propia, la cual resultó ser similarmente eficaz.
Estas metodologías les podrían ayudar a los partidos políticos y a los gobiernos en la toma de decisiones, gracias a entender comportamientos de los diversos sectores de la sociedad que son intrínsecamente anónimos debido al propio proceso democrático, y de esta manera facilitar el direccionamiento de las políticas públicas.


\clearpage
\addtocmark[2]{Author Index} % additional numbered TOC entry
\renewcommand{\indexname}{Author Index}
\clearpage
\addtocmark[2]{Subject Index} % additional numbered TOC entry
\markboth{Subject Index}{Subject Index}
\renewcommand{\indexname}{Subject Index}
\end{document}